\documentclass[a4paper,12pt]{article}

\usepackage{amsmath}
\usepackage[final]{graphicx}
\usepackage{bm}
\usepackage{amssymb}
\usepackage{amssymb}
\newcounter{comment}

{\refstepcounter{comment}%
\begin{quote}
\ttfamily\small$\blacksquare$ \textbf{\underline{Comment} $\sharp$\thecomment:}}%
{\end{quote}}

\begin{document}
\hfill

\begin{center}
\baselineskip=2\baselineskip
\textbf{\LARGE{TeV-scale Seesaw\\
with Quintuplet Fermions}}\\[6ex]
\baselineskip=0.5\baselineskip

{\large Kre\v{s}imir~Kumeri\v{c}ki,
Ivica~Picek
and
Branimir~Radov\v{c}i\'c
}\\[4ex]
\begin{flushleft}
\it
Department of Physics, Faculty of Science, University of Zagreb,
 P.O.B. 331, HR-10002 Zagreb, Croatia\\[3ex]
\end{flushleft}
\today \\[5ex]
\end{center}

\begin{abstract}

We propose a new seesaw model based on a fermionic hypercharge zero weak quintuplet in conjunction with an additional scalar quadruplet which attains an induced vacuum expectation value. The model provides both tree-level seesaw $\sim v^6/M^5$ and loop-suppressed radiative $\sim (1 / 16 \pi^2) \cdot v^2/M$ contributions to active neutrino masses.  The  empirical masses $m_{\nu} \sim 10^{-1}$ eV can be achieved with $M \sim$ TeV new states, accessible at the LHC. For $5\ \rm{fb^{-1}}$ of accumulated
integrated luminosity at the LHC, there could be  $\sim 500$ doubly-charged $\Sigma^{++}$ or $\overline{\Sigma^{++}}$ fermions with mass $M_\Sigma=\rm{400\ GeV}$, leading to interesting multilepton signatures. The neutral component of the fermion quintuplet, previously identified as a minimal dark matter candidate, becomes unstable in the proposed seesaw setup. The stability can be restored by introducing a $Z_2$ symmetry, in which case neutrinos get mass only from radiative contributions.

\end{abstract}
\vspace*{2 ex}

\begin{flushleft}
\end{flushleft}

\clearpage

\section{Introduction}

We propose a seesaw model built upon the hypercharge zero fermionic weak quintuplet  which, in isolation, provides a viable
dark matter (DM) particle within the so-called minimal dark matter (MDM) model \cite{Cirelli:2005uq}. We explore the conditions under which the quintuplets $\Sigma_R \sim (1, 5, 0)$ could simultaneously generate the masses of the  known neutrinos and provide a stable DM candidate.

The proposed model goes beyond the three tree-level realizations
of Weinberg's effective dimension-five operator $LLHH$ \cite{Weinberg:1979sa}: type I
\cite{Minkowski:1977sc-etc}, type II \cite{KoK-etc}, and type III
\cite{Foot:1988aq} seesaw mechanisms, mediated by a heavy fermion singlet,
a scalar triplet, and a fermion triplet, respectively. A seesaw mediator of isospin larger than one has to be accompanied by a scalar multiplet larger than that of the standard model (SM) Higgs doublet.

Our previous model \cite{Picek:2009is,Kumericki:2011hf} based on an exotic nonzero hypercharge weak quintuplet Dirac fermion
$\Sigma_R \sim (1, 5, 2)$ generates the neutrino masses both at the tree level and at the loop level. At the tree level it corresponds to dimension-nine seesaw operator which reproduces the empirical neutrino masses $m_{\nu} \sim 10^{-1}$ eV with $M_\Sigma \sim$ few 100 GeV states testable at the LHC, while for heavier  masses, which are outside of the LHC reach, these exotic leptons generate radiative neutrino masses. However, if one expects from new states to accomplish simultaneously the DM mission, they should have zero hypercharge.

A recent ``R$\nu$MDM model'' \cite{Cai:2011qr} of purely radiative neutrino ($\rm{R\nu}$) masses employs hypercharge zero quintuplets $\Sigma_R \sim (1, 5, 0)$  which, taken in isolation, provide a viable MDM candidate \cite{Cirelli:2005uq}. In a separate paper \cite{KPR12_ScalarDM} we identify an omitted term in \cite{Cai:2011qr} which spoils the claimed DM stability. Therefore, we adopt here a more general approach
in which we analyze in detail both the tree-level and the radiative
neutrino masses generated by hypercharge zero quintuplets
$\Sigma_R \sim (1, 5, 0)$. Thereby we address also the phenomenology of these states at the LHC and recall that by imposing a $Z_2$ symmetry, one can reconcile the seesaw mission with the viability as a DM.

\section{The Model with fermionic quintuplets}

As announced above, the model we propose here is  based on the symmetry of the SM gauge group $SU(3)_C \times SU(2)_L \times U(1)_Y$. In addition to usual SM fermions, we introduce three generations of hypercharge zero quintuplets  $\Sigma_R=(\Sigma_R^{++},\Sigma_R^{+},\Sigma_R^{0},\Sigma_R^{-},\Sigma_R^{--})$, transforming as
$(1,5,0)$ under the gauge group. Also, in addition to the SM Higgs doublet $H= (H^+, H^0)$ there is a scalar quadruplet $\Phi=(\Phi^{+},\Phi^{0},\Phi^{-},\Phi^{--})$ transforming as $(1,4,-1)$.

This is in contrast to a recent model for radiative neutrino masses \cite{Cai:2011qr} adopted subsequently in  \cite{Chen:2011bc}, where a use of the scalar sextuplet $\Phi \sim (1, 6,-1)$ avoids the tree-level contribution. In a tensor notation suitable to cope with higher $SU(2)_L$ multiplets \cite{Cai:2011qr} our additional fields are totally symmetric tensors $\Sigma_{Rijkl}$ and $\Phi_{ijk}$ with the following components:
\begin{eqnarray}
\nonumber
  \Sigma_{R1111} &=& \Sigma_R^{++}\ \ ,\ \ \Sigma_{R1112} = {1\over \sqrt{4}}\Sigma_R^{+}\ \ ,\ \
\Sigma_{R1122} = {1\over \sqrt{6}}\Sigma_R^{0}\ \ , \\
  \Sigma_{R1222} &=& {1\over \sqrt{4}}\Sigma_R^{-}\ \ ,\ \ \Sigma_{R2222} = \Sigma_R^{--}\ \ ;
\end{eqnarray}
\begin{equation}
    \Phi_{111} =  \Phi^{+} \ \ , \ \ \Phi_{112} =  {1\over \sqrt{3}}\Phi^{0} \ \ , \ \ \Phi_{122}
=  {1\over \sqrt{3}}\Phi^{-} \ \ , \ \ \Phi_{222} =  \Phi^{--} \ \ .
\end{equation}
The gauge invariant and renormalizable Lagrangian involving these new fields reads
\begin{equation}\label{lagrangian}
   \mathcal{L} = \overline{\Sigma_R} i \gamma^\mu D_\mu \Sigma_R + (D^\mu \Phi)^\dag (D_\mu \Phi) -
   \big(\overline{L_L} Y \Phi  \Sigma_R + {1 \over 2} \overline{(\Sigma_R)^C} M \Sigma_R + \mathrm{H.c.}\big)
- V(H,\Phi)  \ .
\end{equation}
Here, $D_\mu$ is the gauge covariant derivative, $Y$ is the Yukawa-coupling matrix and $M$ is the mass matrix of the heavy leptons, which we choose to be real and diagonal. For simplicity we drop the flavor indices altogether.

The scalar potential has the gauge invariant form
\begin{eqnarray}\label{scalarpot}
\nonumber  V(H,\Phi) &=& -\mu_H^2 H^\dagger H + \mu_\Phi^2 \Phi^\dagger \Phi + \lambda_1 \big( H^\dagger H \big)^2 + \lambda_2 H^\dagger H \Phi^\dagger \Phi + \lambda_3 H^* H \Phi^* \Phi \\
\nonumber   &+& \big( \lambda_4 H^* H H \Phi+ \mathrm{H.c.} \big) + \big( \lambda_5 H H \Phi \Phi + \mathrm{H.c.} \big) + \big( \lambda_6 H \Phi^* \Phi \Phi+ \mathrm{H.c.} \big)\\
            &+& \lambda_7 \big( \Phi^\dagger \Phi \big)^2 + \lambda_8 \Phi^* \Phi \Phi^* \Phi  \ .
\end{eqnarray}
In the adopted tensor notation the terms in Eqs.~(\ref{lagrangian}) and ~(\ref{scalarpot}) read
\begin{eqnarray}\label{tensor}
\overline{L_L} \Phi \Sigma_R = \overline{L_L}^i \Phi_{jkl} \Sigma_{Rij'k'l'} \epsilon^{jj'}\epsilon^{k k'}\epsilon^{l l'} &,& \overline{(\Sigma_R)^C} \Sigma_R = \overline{(\Sigma_R)^C}_{ijkl} \Sigma_{Ri'j'k'l'}
\epsilon^{ii'}\epsilon^{jj'}\epsilon^{kk'}\epsilon^{ll'}\ ,\nonumber\\
H^* H \Phi^* \Phi = H^{*i} H_j \Phi^{*jkl} \Phi_{ikl} &,& H^* H H \Phi = H^{*i} H_j H_k \Phi_{ij'k'} \epsilon^{jj'}\epsilon^{kk'}   \ ,\nonumber\\
H H \Phi \Phi =H_i \Phi_{jkl} H_{i'} \Phi_{j'k'l'} \epsilon^{ij}\epsilon^{i'j'}\epsilon^{kk'}\epsilon^{ll'}  &,& H \Phi^* \Phi \Phi = H_i \Phi^{*ijk} \Phi_{jlm} \Phi_{kl'm'} \epsilon^{ll'}\epsilon^{mm'} \ ,\nonumber\\
\Phi^* \Phi \Phi^* \Phi = \Phi^{*ijk} \Phi_{i'jk} \Phi^{*i'j'k'} \Phi_{ij'k'} &.&
\end{eqnarray}
Accordingly, the Majorana mass term for the quintuplet $\Sigma_R$ is expanded in component fields to give
\begin{eqnarray}
\nonumber   \overline{(\Sigma_R)^C} M \Sigma_R &=& \overline{(\Sigma^{++}_R)^C} M \Sigma_R^{--} - \overline{(\Sigma^+_R)^C} M \Sigma_R^{-} + \overline{(\Sigma^0_R)^C} M \Sigma_R^0 \\
   &-& \overline{(\Sigma^-_R)^C} M \Sigma_R^{+} + \overline{(\Sigma^{--}_R)^C} M \Sigma_R^{++}\ ,
\end{eqnarray}
the terms containing two charged Dirac fermions and one neutral Majorana fermion
\begin{equation}\label{fermions}
   \Sigma^{++} = \Sigma^{++}_R + \Sigma^{--C}_R\ ,\ \Sigma^+ = \Sigma^+_R - \Sigma^{-C}_R\ ,\ \Sigma^0
= \Sigma^0_R + \Sigma^{0C}_R\ .
\end{equation}
The electroweak symmetry breaking proceeds in usual way from the vacuum expectation value (vev) $v$ of the Higgs doublet, corresponding to the negative sign in front of the $\mu_H^2$ term in Eq.~(\ref{scalarpot}). On the other hand, the electroweak $\rho$ parameter dictates a small value for vev $v_\Phi$ of the scalar quadruplet, implying the positive sign in front
of the $\mu_\Phi^2$ term. However, the presence of the $\lambda_4$ term in  Eq.~(\ref{scalarpot}), given explicitly by
\begin{eqnarray}
\nonumber
H^* H H \Phi = \frac{1}{\sqrt{3}} H^{+*} H^+ H^+ \Phi^- - \frac{2}{\sqrt{3}} H^{0*} H^+ H^0 \Phi^- + H^{0*} H^+ H^+ \Phi^{--} \\
+  H^{+*} H^0 H^0 \Phi^+ - \frac{2}{\sqrt{3}} H^{+*} H^+ H^0 \Phi^0 + \frac{1}{\sqrt{3}} H^{0*} H^0 H^0 \Phi^0  \ ,
\end{eqnarray}
leads to the induced vev for the $\Phi^0$ field,
\begin{equation}\label{phivev}
    v_\Phi \simeq - \frac{1}{\sqrt{3}} \lambda_4^* \frac{v^3}{\mu_\Phi^2} \, \, .
\end{equation}
Since it changes the electroweak $\rho$ parameter from the unit value to $ \rho\simeq1+6 v_\Phi^2/v^2$, a comparison to the experimental value $\rho=1.0008^{+0.0017}_{-0.0007}$ \cite{PDG10} gives us a constraint $v_\Phi \lesssim 3.6$ GeV.

\section{Neutrino masses}

\begin{figure}
\centerline{\includegraphics[scale=1.30]{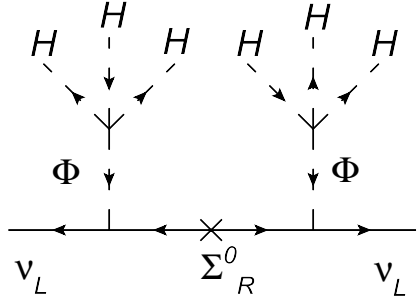}}
\caption{\small Tree-level diagram corresponding to dimension-nine operator in Eq.~(\ref{tree}). The fermion line flow indicates a Majorana nature of the seesaw mediator.}
\label{dim9op}
\end{figure}

The Yukawa interaction terms from Eq.~(\ref{lagrangian}), when expressed like that in  Eq.~(\ref{tensor}), read explicitly as
\begin{eqnarray}
\nonumber
\overline{L_L} \Phi \Sigma_R = -\frac{\sqrt{3}}{2} \bar{\nu }_L \Phi^- \Sigma _R^+ - \frac{1}{\sqrt{2}} \bar{l}_L \Phi^- \Sigma_R^0
+ \frac{\sqrt{3}}{2} \bar{l}_L \Phi^0 \Sigma _R^- + \frac{1}{\sqrt{2}} \bar{\nu }_L \Phi^0 \Sigma_R^0 \\
- \frac{1}{2} \bar{\nu }_L \Phi^+ \Sigma _R^- - \bar{l}_L \Phi^+ \Sigma _R^{--}
+ \frac{1}{2} \bar{l}_L \Phi^{--} \Sigma _R^+  + \bar{\nu }_L \Phi^{--} \Sigma _R^{++}  \, .
\end{eqnarray}
The vev $v_\Phi$ generates a Dirac mass term connecting $\nu_L$ and $\Sigma_R^0$, a nondiagonal entry in
the mass matrix for neutral leptons given by
\begin{eqnarray}
\mathcal{L}_{\nu \Sigma^0} =  \, -\frac{1}{2}
\left(  \overline{\nu_L}  \; \overline{(\Sigma_R^0)^C} \right)
\left( \! \begin{array}{cc}
0 & \frac{1}{\sqrt{2}} Y v_\Phi \\
\frac{1}{\sqrt{2}} Y^T v_\Phi & M
\end{array} \! \right) \,
\left( \!\! \begin{array}{c} (\nu_L)^c \\ \Sigma_R^0 \end{array} \!\! \right)
\; + \mathrm{H.c.}\ .
\label{neutral_mass_matrix}
\end{eqnarray}
A similar term connecting light and heavy charged leptons  gives the mass matrix for charged leptons
\begin{eqnarray}
\mathcal{L}_{l \Sigma^-} =  \, -
\left(  \overline{l_L}  \; \overline{(\Sigma_R^+)^C} \right)
\left( \! \begin{array}{cc}
Y_l v & -\frac{\sqrt{3}}{2} Y v_\Phi \\
0 & M
\end{array} \! \right) \,
\left( \!\! \begin{array}{c} l_R \\ (\Sigma_L^+)^C \end{array} \!\! \right)
\; + \mathrm{H.c.}\ .
\label{charged_mass_matrix}
\end{eqnarray}
After diagonalizing the mass matrix for the neutral leptons, the light neutrinos acquire the Majorana mass matrix given by
\begin{equation}
    m_\nu^{tree} = - \frac{1}{2} v_\Phi^2 Y M^{-1} Y^T \ .
\end{equation}
In the basis where the matrix of heavy leptons is real and diagonal, $M= diag (M_1,M_2,M_3)$, we can write $m_\nu^{tree}$ as
\begin{equation}
    (m_\nu)_{ij}^{tree} = - \frac{1}{2} v_\Phi^2 \sum_k {Y_{ik} Y_{jk} \over M_k} \ .
\end{equation}
Together with the induced  vev in Eq.~(\ref{phivev}), this gives
\begin{equation}\label{tree}
    (m_{\nu})_{ij}^{tree} = - \frac{1}{6} (\lambda^*_4)^2 \frac{v^6}{\mu_\Phi^4} \sum_k {Y_{ik} Y_{jk} \over M_k} \,,
\end{equation}
which reflects the fact that the light neutrino mass is generated from the dimension-nine operator corresponding to the tree-level seesaw mechanism displayed on Fig.~\ref{dim9op}.

\begin{figure}
\centerline{\includegraphics[scale=1.30]{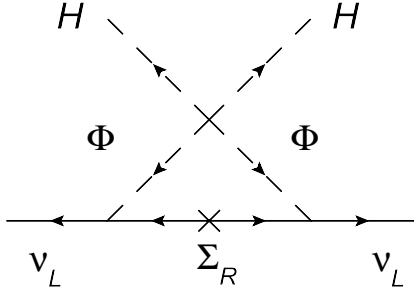}}
\caption{\small One-loop diagrams generating the light neutrino masses.}
\label{oneloop}
\end{figure}

Besides at the tree level, the light neutrino masses arise also through one-loop diagrams displayed on Fig.~\ref{oneloop}. The crucial quartic $\lambda_5$ term in Eq.~(\ref{scalarpot}), when expanded in component fields, gives
\begin{eqnarray}
\nonumber
H H \Phi \Phi = \frac{2}{\sqrt{3}} \Phi^- \Phi^+ H^0 H^0 - \frac{2}{3} \Phi^- \Phi^- H^+ H^+ + \frac{2}{\sqrt{3}} \Phi^{--} \Phi^0 H^+ H^+ \\
- \frac{2}{3} \Phi^0 \Phi^0 H^0 H^0 + \frac{2}{3} \Phi^- \Phi^0 H^+ H^0 - 2 \Phi^{+} \Phi^{--} H^+ H^0  \,.
\end{eqnarray}

\begin{figure}
\centerline{\includegraphics[scale=1.00]{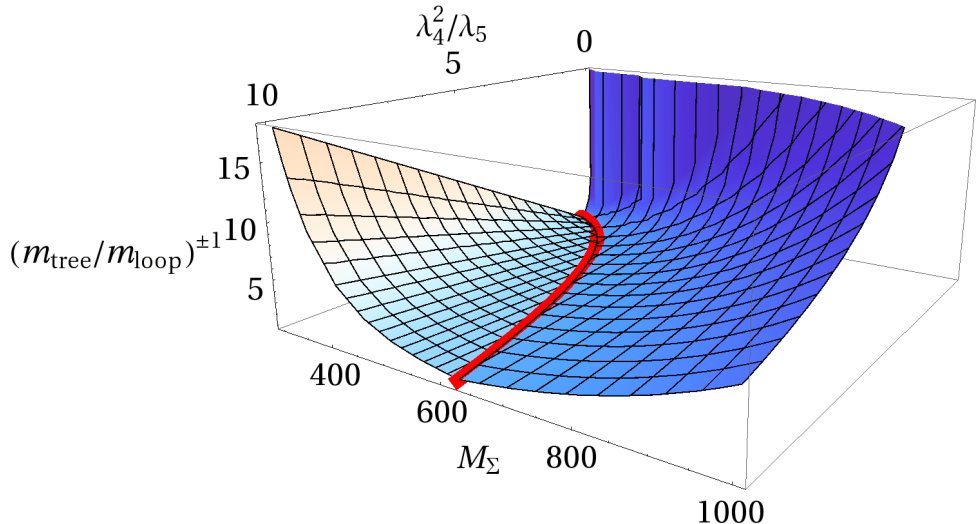}}
\caption{\small For the portion of the parameter space left from the equality-division line, where the tree-level contribution to neutrino masses dominates, we plot $m_{tree}/m_{loop}$. Right from the equality-division line the one-loop contribution to neutrino masses dominates, and we plot $m_{loop}/m_{tree}$.}
\label{tree_dom-ce}
\end{figure}

There are two different contributions on Fig.~\ref{oneloop}, one with heavy neutral fields ($\Sigma^0,\Phi^0$) and the other with heavy charged fields ($\Sigma^+,\Phi^+,\Phi^-$) running in the loop. If we neglect the mass splitting within the $\Sigma$ and $\Phi$ multiplets, the contribution to the light neutrino mass matrix is given by
\begin{equation}\label{loop}
(m_\nu)_{ij}^{loop} = {-5 \lambda_5^* v^2 \over 24 \pi^{2}}
\sum_k {Y_{ik} Y_{jk} M_k \over m_\Phi^{2} - M_k^{2}} \left[
1 - {M_k^{2} \over m_\Phi^{2}-M_k^{2}} \ln {m_\Phi^{2} \over M_k^{2}}  \right] \,.
\end{equation}
This expression can be further specified \cite{Ma:2006km}, depending on relative values of the masses for additional fermion and scalar multiplets. For heavy fermions significantly heavier than the new scalars, $M_k^{2} >> m_\Phi^{2}$,
\begin{equation}
(m_\nu)_{ij}^{loop} = {-5 \lambda_5^* v^2 \over 24 \pi^{2}}
\sum_k {Y_{ik} Y_{jk} \over M_k} \left[
\ln {M_k^{2} \over m_\Phi^{2}} - 1 \right] \,.
\end{equation}
In the opposite case, for $m_\Phi^{2} >> M_k^{2}$,
\begin{equation}
(m_\nu)_{ij}^{loop} = {-5 \lambda_5^* v^2 \over 24 \pi^{2}m_\Phi^{2}}
\sum_k Y_{ik} Y_{jk} M_k   \,.
\end{equation}
Finally, if $m_\Phi^{2} \simeq M_k^{2}$, then
\begin{equation}\label{simeq}
(m_\nu)_{ij}^{loop} = {-5 \lambda_5^* v^2 \over 48 \pi^{2}}
\sum_k {Y_{ik} Y_{jk} \over M_k}.
\end{equation}
The tree-level and the loop contributions added together give the light neutrino mass matrix
\begin{eqnarray}\label{mnu}
  (m_\nu)_{ij} &=& (m_\nu)_{ij}^{tree}+(m_\nu)_{ij}^{loop} \nonumber\\
               &=& \frac{-1}{6} (\lambda^*_4)^2 \frac{v^6}{\mu_\Phi^4} \sum_k {Y_{ik} Y_{jk} \over M_k}
+ {-5 \lambda_5^* v^2 \over 24 \pi^{2}}
\sum_k {Y_{ik} Y_{jk} M_k \over m_\Phi^{2} - M_k^{2}} \left[
1 - {M_k^{2} \over m_\Phi^{2}-M_k^{2}} \ln {m_\Phi^{2} \over M_k^{2}}  \right] \,. \nonumber\\
\end{eqnarray}
In the case of comparable masses the light neutrino mass matrix reads
\begin{equation}\label{mnuapp}
    (m_\nu)_{ij} =  \left[ \frac{-1}{6} (\lambda^*_4)^2 \frac{v^6}{\mu_\Phi^4} +
{-5 \lambda_5^* v^2 \over 48 \pi^{2}}\right] \sum_k {Y_{ik} Y_{jk} \over M_k} \,.
\end{equation}
From this expression we can estimate the high energy scale of our model and the corresponding value for $v_\Phi$ in Eq.~(\ref{phivev}). For illustrative purposes we take the same values for the mass parameters ($\mu_\Phi=M_\Sigma=\Lambda_{NP}$) and the empirical input values $v=174$ GeV and $m_\nu\sim0.1$ eV. For the moderate values $Y \sim 10^{-3}\ ,\ \lambda_4 \sim 10^{-2}$ and $\lambda_5 \sim 10^{-4}$ we get $\Lambda_{NP}\simeq 440$ GeV and $v_\Phi\simeq 160$ MeV.
Corresponding masses of the new states could be tested at the LHC.
On Fig.~\ref{tree_dom-ce} we display the part of the parameter space for which the tree-level (loop-level) contribution
dominates.

\section{Production of quintuplet leptons at the LHC}
\label{sec:production}

The production channels of the  heavy quintuplet leptons in proton-proton collisions are
dominated by the quark-antiquark annihilation via neutral and charged gauge bosons
\begin{displaymath}
q + \bar{q} \to A \to \Sigma + \bar{\Sigma}\;, \qquad A = \gamma, Z, W^\pm \;,
\end{displaymath}
where the gauge Lagrangian relevant for the production is given by
\begin{eqnarray}\label{gauge-short}
\mathcal{L}_{gauge}^{\Sigma \overline{\Sigma}}=
&+& e (2\overline{\Sigma^{++}}\gamma^\mu \Sigma^{++} + \overline{\Sigma^{+}}\gamma^\mu \Sigma^{+})A_\mu \nonumber \\
&+& g  \cos\theta_W( 2 \overline{\Sigma^{++}}\gamma^\mu \Sigma^{++} + \overline{\Sigma^{+}}\gamma^\mu\Sigma^{+})Z_\mu\\
&+&g(\sqrt{2} \overline{\Sigma^{+}}\gamma^\mu\Sigma^{++} + \sqrt{3}\overline{\Sigma^{0}}\gamma^\mu\Sigma^{+}) W^-_\mu + \mathrm{H.c.}\ .\nonumber
\end{eqnarray}
The cross section for the partonic process is
\begin{equation}
\hat{\sigma} (q \bar{q} \to \Sigma \overline{\Sigma}) =
\frac{\beta (3-\beta^2)}{48 \pi} \hat{s} (V_{L}^2 + V_{R}^2) \;,
\label{eq:xspartonic}
\end{equation}
where $\hat{s}\equiv(p_q + p_{\bar{q}})^{2}$ is the Mandelstam variable $s$ for the
quark-antiquark system, the parameter $\beta\equiv\sqrt{1-4 M_\Sigma^2/\hat{s}}$ denotes the
heavy lepton velocity, and the left- and right-handed couplings are given by
\begin{align}
V_{L,R}^{(\gamma+Z)}&= \frac{Q_{\Sigma} Q_q e^2}{\hat{s}} +
    \frac{g^{Z\Sigma}\, g_{L,R}^{q} \, g^2}{c_{W}^2(\hat{s}-M_{Z}^2)}\;, \\
V_{L}^{(W^-)}&= \frac{g^{W\Sigma} g^2 V_{ud}}{\sqrt{2}(\hat{s}-M_{W}^2)} =
  V_{L}^{(W^+)^*}\;,  \\
V_{R}^{(W^\pm)}&= 0 \;.
\end{align}
Here,
$g^{q}_L = T_3 - s_{W}^2 Q_q $ and $g^{q}_R = - s_{W}^2 Q_q$
are the SM chiral quark couplings to the $Z$ boson.
The vector couplings of heavy leptons to gauge bosons are
\begin{equation}\label{vectorcouplings}
g^{Z\Sigma} = T_3 - s_{W}^2 Q_\Sigma\ \rm{and} \ g^{W\Sigma} = \sqrt{2}\ \rm{or}\ \sqrt{3}\ ,
\end{equation}
where $g^{W\Sigma}$ can be read of the last raw in Eq.~(\ref{gauge-short}),
relevant for the production of
$\Sigma^{++}\overline{\Sigma^{+}}$ and
$\Sigma^{+}\overline{\Sigma^{0}}$ pairs, respectively.

In evaluating the cross sections for a hadron collider, the partonic cross section
(\ref{eq:xspartonic}) has to be convoluted with the appropriate parton distribution
functions (PDFs). To evaluate the cross sections we have used CTEQ6.6 PDFs \cite{Nadolsky:2008zw}
via LHAPDF software library \cite{Whalley:2005nh}.

\begin{figure}
\begin{center}
\includegraphics[scale=0.70]{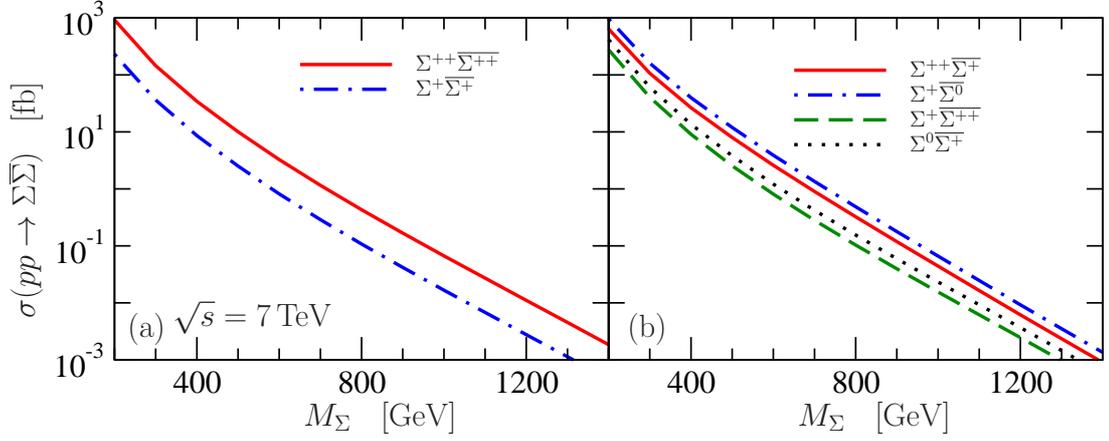}
\end{center}
\caption{The cross sections for production of quintuplet lepton  pairs on LHC
proton-proton collisions at $\sqrt{s}=7\, {\rm TeV}$ via
neutral $\gamma, Z$ (a) and charged $W^{\pm}$ currents (b),
in dependence on the heavy quintuplet mass $M_{\Sigma}$.}
\label{QuiY0}
\end{figure}

\begin{figure}
\begin{center}
\includegraphics[scale=0.70]{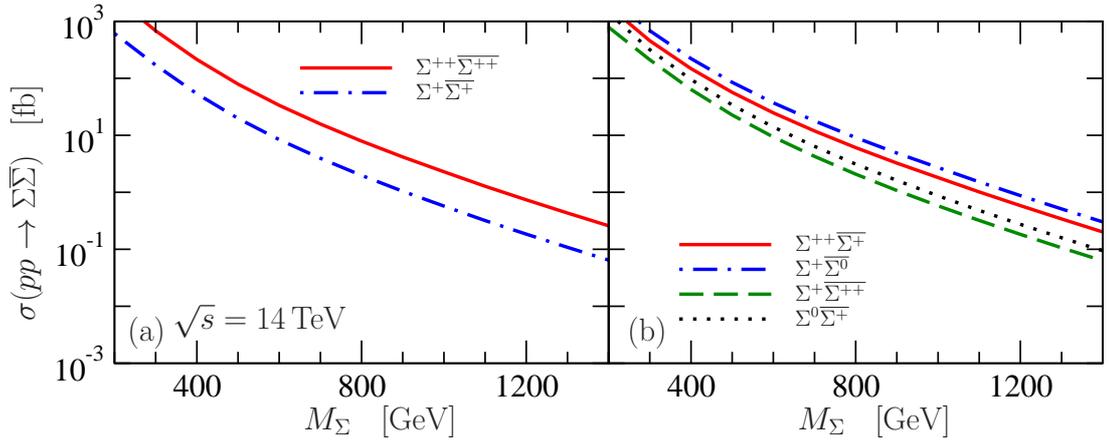}
\end{center}
\caption{Same as Fig.~\ref{QuiY0}, but for designed  $\sqrt{s}=14\, {\rm TeV}$ at the LHC.}
\label{QuiY014}
\end{figure}

The cross sections for proton-proton collisions are presented on Fig.~\ref{QuiY0} for $\sqrt{s} = 7\, {\rm TeV}$
appropriate for the 2011 LHC run, and for designed $\sqrt{s}=14\,{\rm TeV}$ on Fig.~\ref{QuiY014}. Thereby we distinguish separately the production via neutral currents shown on the left panel, and via charged currents shown on the right panel of Figs.~\ref{QuiY0} and \ref{QuiY014}.


\begin{table}
\begin{center}
\begin{tabular}{cccc}
\hline
Produced   &                & Cross section ($\rm{fb}$)  &              \\
pair      & $M_\Sigma = 200\ \rm{GeV}$  & $M_\Sigma = 400\ \rm{GeV}$  &  $M_\Sigma = 800\ \rm{GeV}$ \\ \hline
$\Sigma^{++} \overline{\Sigma^{++}}$     & 924    &  34.4   & 0.43 \\
$\Sigma^{+} \overline{\Sigma^{+}}$       & 231    &   8.6   & 0.11 \\ \hline

$\Sigma^{++} \overline{\Sigma^{+}}$      & 641     & 26.5   & 0.32  \\
$\Sigma^{+} \overline{\Sigma^{0}}$       & 961     & 39.8   & 0.49  \\ \hline

$\Sigma^{+} \overline{\Sigma^{++}}$      & 276  &  9.1  & 0.10  \\
$\Sigma^{0} \overline{\Sigma^{+}}$       & 414  & 13.6  & 0.15  \\ \hline \hline
Total     & 3447   & 132    & 1.6  \\ \hline
\end{tabular}
\caption{Production cross sections for $\Sigma$-$\overline{\Sigma}$ pairs for the
LHC run at $\sqrt{s} = 7\, {\rm TeV}$, for three selected values of $M_\Sigma$.}
\label{table}
\end{center}
\end{table}

\begin{figure}
\begin{center}
\includegraphics[scale=0.8]{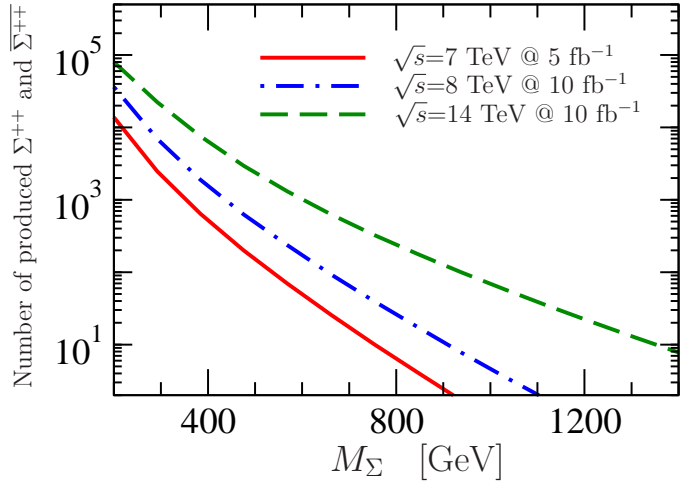}
\end{center}
\caption{Number of $\Sigma^{++}$ and
$\overline{\Sigma^{++}}$ particles produced for three
characteristic LHC collider setups, in dependence on
the heavy lepton mass $M_{\Sigma}$.}
\label{fig:LHCproduction}
\end{figure}

\newpage

We extract in Table \ref{table} the pair production cross sections
from Fig.~\ref{QuiY0} for three selected values of $M_\Sigma$.
From this table we find that the doubly-charged
$\Sigma^{++}$ has the largest production cross section
$\sigma(\Sigma^{++})|_{M_\Sigma = 400 \rm{GeV}}= 60.9\, \rm{fb}$ and doubly-charged $\overline{\Sigma^{++}}$ has the smallest, still comparable $\sigma(\overline{\Sigma^{++}})|_{M_\Sigma = 400 \rm{GeV}}= 43.5\, \rm{fb}$. The production rates of other heavy leptons are in between. On Fig. \ref{fig:LHCproduction} we
plot the expected number of produced $\Sigma^{++}$ and
$\overline{\Sigma^{++}}$ particles for three
characteristic collider setups.
In particular, for $5\ \rm{fb^{-1}}$ of integrated luminosity
of 2011 LHC run at $\sqrt{s}=7$ TeV, there should be
about 520 doubly-charged $\Sigma^{++}$ or
$\overline{\Sigma^{++}}$ fermions produced. In total, there should be 660 $\Sigma - \overline{\Sigma}$ pairs produced.

By testing the heavy lepton production cross sections one can hope to identify the quantum numbers of the quintuplet particles,
but in order to confirm their relation to neutrinos one has to study their decays.

\section{Decays of quintuplet states}
\label{sec:triplychargeddecay}

Our focus here is entirely on the decay modes of the heavy lepton states listed in Eq.~(\ref{fermions}).
Namely, provided that in our scenario the exotic scalar states are slightly heavier than the exotic leptons, the exotic scalar fields will not appear in the final states of  heavy lepton decays. Note that exotic scalar states $\Phi \sim (1, 4, 1)$ were considered recently in \cite{Ren2011}, in the context of modified type III seesaw model.

The couplings relevant for the decays at hand stem from off-diagonal entries in the mass matrices in Eqs.~(\ref{neutral_mass_matrix}) and (\ref{charged_mass_matrix}). The diagonalization of these matrices can be achieved by making the following unitary transformations on the lepton fields:
\begin{equation}\nonumber
    \left(
    \begin{array}{c}
      (\nu_{L})^c \\
      \Sigma_R^0 \\
    \end{array}
  \right)
  =U^0
  \left(
    \begin{array}{c}
      (\nu_{mL})^c \\
      \Sigma_{mR}^0 \\
    \end{array}
  \right)\ ,
\end{equation}
\begin{equation}
    \left(
    \begin{array}{c}
      l_L \\
      (\Sigma_R^+)^c \\
    \end{array}
  \right)
  =U^L
  \left(
    \begin{array}{c}
      l_{mL} \\
      (\Sigma_{mR}^+)^c \\
    \end{array}
  \right)\ , \ \
\left(
    \begin{array}{c}
      l_R \\
      (\Sigma_L^+)^c \\
    \end{array}
  \right)
  =U^R
  \left(
    \begin{array}{c}
      l_{mR} \\
      (\Sigma_{mL}^+)^c \\
    \end{array}
  \right)\ .
\end{equation}
Following the procedure in Refs.~\cite{Kumericki:2011hf, Grimus:2000vj, Li:2009mw}, the matrices $U^{0}$ and $U^{L,R}$ can be expressed in terms of $Y v_\Phi$ and $M$. Thereby, by expanding $U^{0}$ and $U^{L,R}$ in powers of $M^{-1}$ and keeping only the leading order terms, we get
\begin{equation}
U^0\equiv \left(
  \begin{array}{cc}
    V_{PMNS}^* & \frac{1}{\sqrt{2}} v_\Phi^* Y^* M^{-1} \\
    -\frac{1}{\sqrt{2}} v_\Phi M^{-1} Y^T V_{PMNS}^* & 1\\
  \end{array}
\right)\ ,
\end{equation}
\begin{equation}
U^L\equiv \left(
  \begin{array}{cc}
    1 & -\frac{\sqrt{3}}{2} v_\Phi Y M^{-1} \\
    \frac{\sqrt{3}}{2} v_\Phi M^{-1} Y^\dag & 1\\
  \end{array}
\right)\ ,\
U^R\equiv \left(
  \begin{array}{cc}
    1 & 0 \\
    0 & 1\\
  \end{array}
\right)\ .
\end{equation}
Here, $V_{PMNS}$ is a $3 \times 3$ unitary matrix which diagonalizes the effective light neutrino mass matrix in Eq.~(\ref{mnu}),
and the nondiagonal entries are related to the sought couplings of heavy and light leptons, and are expressed  in terms of the matrix-valued quantity
\begin{equation}
    V_{l\Sigma}=  (v_\Phi Y M^{-1})_{l\Sigma}  \, .
\end{equation}
Next, for simplicity, we suppress the indices indicating the mass-eigenstate fields. The Lagrangian in the mass-eigenstate basis, relevant for the decays of the heavy leptons, has the neutral current part
\begin{eqnarray}
  \mathcal{L}_{NCZ} &=& {g\over
c_W}\Big[ \overline{\nu} ( {1 \over 2 \sqrt 2} V_{PMNS}^\dagger V \gamma^\mu P_L
-{1 \over 2 \sqrt 2} V_{PMNS}^T V^* \gamma^\mu P_R ) \Sigma^0 \nonumber \\
&+& \overline{l^c} ({\sqrt 3 \over 4} V^*  \gamma^\mu P_R )
\Sigma^+ + \mathrm{H.c.} \Big] Z_\mu^0\ ,
\end{eqnarray}
and the charged current part
\begin{eqnarray}
\mathcal{L}_{CC} &=& g\Big[\overline{\nu}(- \sqrt {3 \over 2} V_{PMNS}^\dagger V \gamma^\mu P_L + {- \sqrt 3\over 2 \sqrt 2} V_{PMNS}^T V^* \gamma^\mu P_R)\Sigma^+\nonumber\\
&+& \overline{l} (- V \gamma^\mu P_L) \Sigma^0 \nonumber\\
&+& \overline{l^c} (\sqrt{3\over 2} V^* \gamma^\mu P_R) \Sigma^{++} \Big] W_\mu^- + \mathrm{H.c.}\ .
\end{eqnarray}

Let us start our list of the partial decay widths by the decays of neutral $\Sigma^0$ state:
\begin{eqnarray}\label{width0}
&&\Gamma(\Sigma^0\to \ell^\mp W^\pm)= {g^2\over 32\pi}\Big|V_{\ell \Sigma}\Big|^2 {M_\Sigma^3\over M_W^2}\left(1-{M_W^2\over
M_\Sigma^2}\right)^2\left(1+2{M_W^2\over M_\Sigma^2}\right),\nonumber\\
&&\sum^3_{m=1}\Gamma(\Sigma^0\to \nu_m Z^0)= {g^2\over 32\pi c_W^2} \sum_{\ell=\{e, \mu, \tau \}} {1\over 4} \Big|V_{\ell \Sigma}\Big|^2 {M_\Sigma^3\over M_Z^2}\left(1-{M_Z^2\over
M_\Sigma^2}\right)^2\left(1+2{M_Z^2\over M_\Sigma^2}\right).\nonumber\\
\end{eqnarray}
The positive singly-charged heavy lepton $\Sigma^+$ has the following partial decay widths:
\begin{eqnarray}\label{width+}
&&\Gamma(\Sigma^+\to \ell^+Z^0)= {g^2\over 32\pi c_W^2} {3\over 16} \Big|V_{\ell \Sigma}\Big|^2 {M_\Sigma^3\over M_Z^2}\left(1-{M_Z^2\over
M_\Sigma^2}\right)^2\left(1+2{M_Z^2\over M_\Sigma^2}\right),\nonumber\\
&&\sum^3_{m=1}\Gamma(\Sigma^+\to \nu_m W^+)= {g^2\over 32\pi} \sum_{\ell=\{e, \mu, \tau \}} {15\over 8} \Big| V_{\ell \Sigma}\Big|^2 {M_\Sigma^3\over M_W^2}\left(1-{M_W^2\over M_\Sigma^2}\right)^2\left(1+2{M_W^2\over M_\Sigma^2}\right). \nonumber\\
\end{eqnarray}
Finally, the doubly-charged $\Sigma^{++}$ state decays exclusively via a charged current, with the partial decay width
\begin{equation}\label{width++}
    \Gamma(\Sigma^{++}\to \ell^+W^+)= {g^2\over 32\pi} {3\over 2} \Big| V_{\ell \Sigma}\Big|^2 {M_\Sigma^3\over M_W^2}\left(1-{M_W^2\over
M_\Sigma^2}\right)^2\left(1+2{M_W^2\over M_\Sigma^2}\right).
\end{equation}

The mass difference induced by loops of SM gauge bosons between two components of $\Sigma$ quintuplet with electric charges $Q$ and $Q'$ is explicitly calculated in \cite{Cirelli:2005uq}
\begin{eqnarray}\label{massdifferences}
M_Q -  M_{Q'} &=&\frac{\alpha_2 M}{4\pi} \Big\{ (Q^2-Q^{\prime 2})s_{\rm W}^2 f(\frac{M_Z}{M})+(Q-Q')(Q+Q'-Y) \nonumber \\  && \bigg[f(\frac{M_W}{M})-f(\frac{M_Z}{M})\bigg] \Big\}\ , \\
f(r) &=& {r \over 2}  \left[2 r^3\ln r -2 r+(r^2-4)^{1/2} (r^2+2) \ln \left( {r^2 -2 - r\sqrt{r^2-4} \over 2} \right)\right] \ .\nonumber
\end{eqnarray}
The values for the mass splittings $\Delta M_{ij}=M_i-M_j$ in Eq.~(\ref{massdifferences}) for $M_{\Sigma}=400\ \rm{GeV}$ are
\begin{equation}\label{splitting}
\Delta M_{21} \equiv M_{\Sigma^{++}} - M_{\Sigma^{+}} \simeq 490\ \rm{MeV}\ ,\ \Delta M_{10} \equiv M_{\Sigma^{+}} - M_{\Sigma^{0}} \simeq 163\ \rm{MeV}\ ,
\end{equation}
which opens additional decay channels, like $\Sigma^{++} \to \pi^+ \Sigma^+$ and $\Sigma^{+} \to \pi^+ \Sigma^0$.
The decay rate for a single pion finale state is given by
\begin{equation}\label{widthpion}
\Gamma(\Sigma^i \to \Sigma^j \pi^+ )=  (g^{W\Sigma})^2_{ij} {2 \over \pi}  G_{\rm F}^2 |V_{ud}|^2 f_\pi^2 (\Delta M_{ij})^3
\sqrt{1-\frac{m_\pi^2}{(\Delta M_{ij})^2}}
\end{equation}
where $(g^{W\Sigma})^2_{ij}$ is given in Eq.~(\ref{vectorcouplings}). These  decays are suppressed by small mass differences.
\begin{figure}
\begin{center}
\includegraphics[scale=0.70]{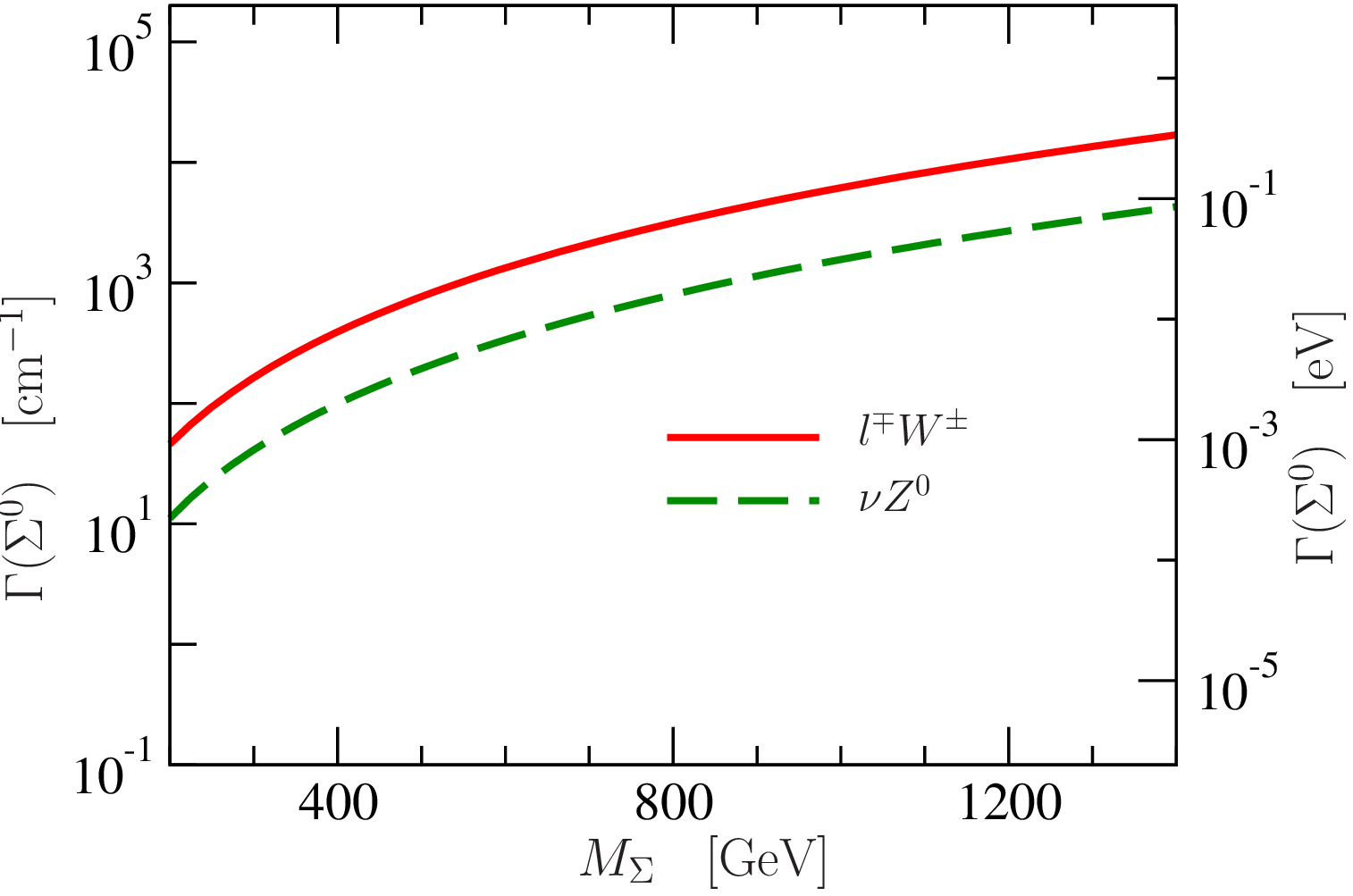}
\end{center}
\caption{Partial decay widths of $\Sigma^0$
quintuplet lepton for $|V_{l\Sigma}| = 3.5 \cdot 10^{-7}$, in dependence
on heavy quintuplet mass $M_{\Sigma}$.}
\label{decay0}
\end{figure}

\begin{figure}
\begin{center}
\includegraphics[scale=0.70]{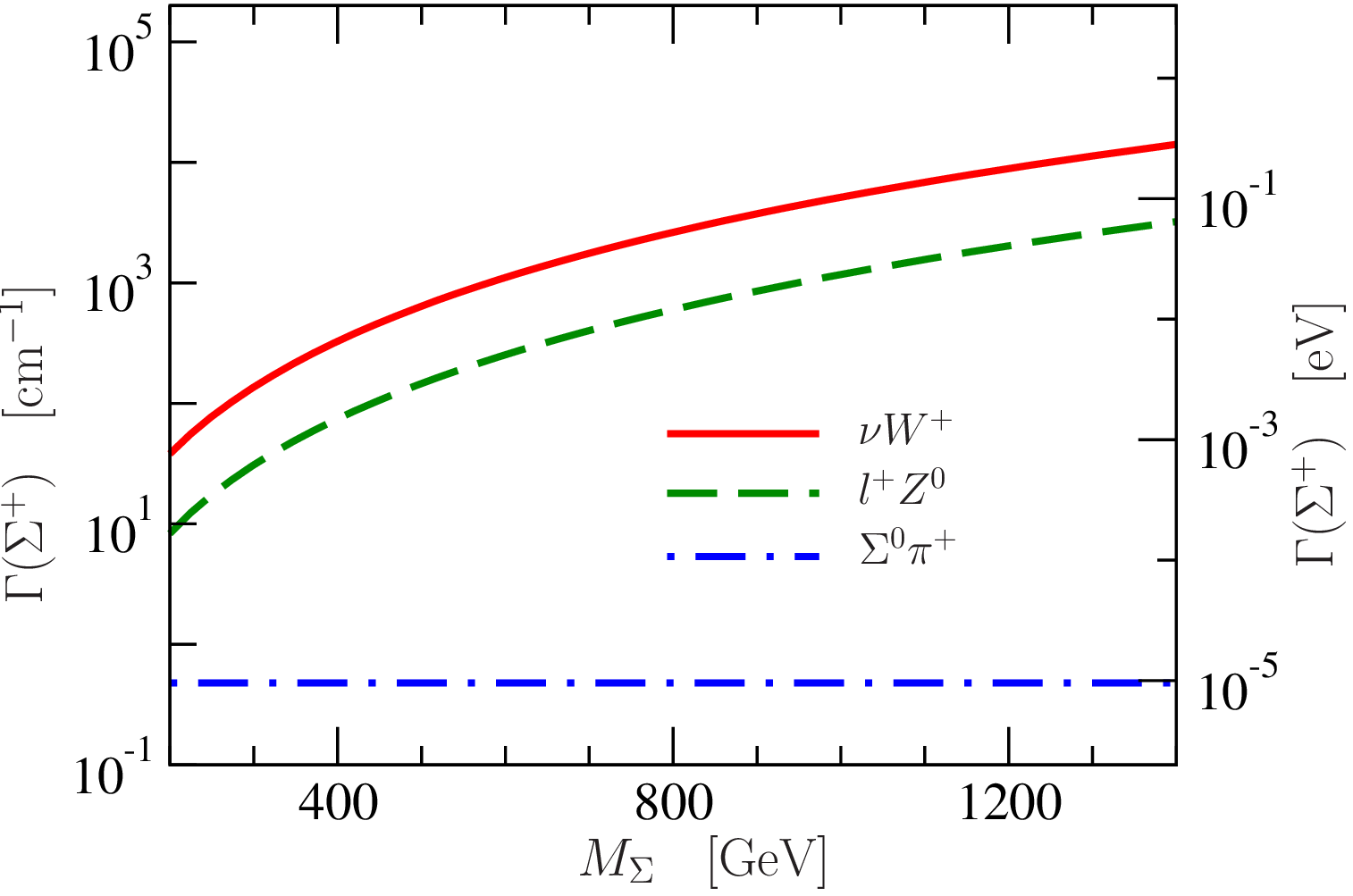}
\end{center}
\caption{Partial decay widths of $\Sigma^+$
quintuplet lepton for $|V_{l\Sigma}| = 3.5 \cdot 10^{-7}$, in dependence
on heavy quintuplet mass $M_{\Sigma}$.}
\label{decay+}
\end{figure}
\begin{figure}
\begin{center}
\includegraphics[scale=0.66]{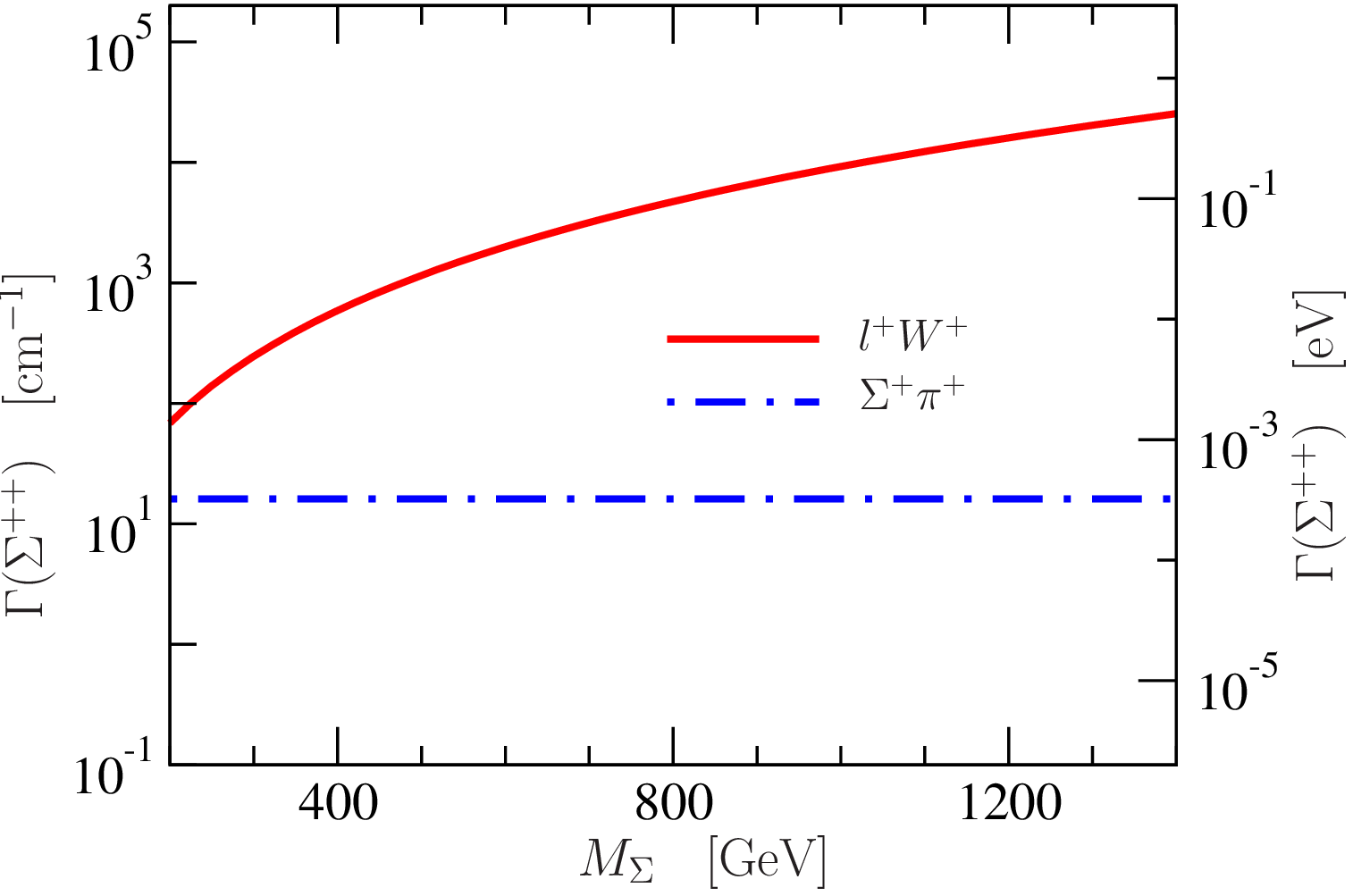}
\end{center}
\caption{Partial decay widths of $\Sigma^{++}$
quintuplet lepton for $|V_{l\Sigma}| = 3.5 \cdot 10^{-7}$, in dependence
on heavy quintuplet mass $M_{\Sigma}$.}
\label{decay++}
\end{figure}

In Figs.~\ref{decay0},\ref{decay+}, and \ref{decay++} we plot the partial widths of the decays of $\Sigma^0$, $\Sigma^+$ and $\Sigma^{++}$ given in Eqs.~(\ref{width0}), (\ref{width+}) and (\ref{width++}), respectively.  Figures.~\ref{decay+} and \ref{decay++} also show pion final state decays from Eq.~(\ref{widthpion}). We list the representative final states of these decays in Table~\ref{events}, which includes same-sign dilepton events as distinguished signatures at the LHC.

\begin{table}[h]
\centering
\begin{tabular}{|c||c|c|c|c|} \hline
 &  $\overline{\Sigma^{++}}\to \ell^-W^- $   &
 $\overline{\Sigma^+}\to \ell^-Z^0 $   &   $\Sigma^0\to \ell^+W^- $   &
 $\Sigma^0\to \ell^-W^+ $ \\
 &  $(0.66)$   &
 $(0.06)$   &   $(0.30)$   &
 $(0.30)$ \\\hline
\hline
$\Sigma^{++}\to \ell^+W^+ $   &  $\ell^+ \ell^-W^+ W^- $   &
 $\ell^+ \ell^- W^+ Z^0 $     &   -     &  -  \\
$ (0.66)$   &  $ (0.44)$   &
 $ (0.04)$     &   -     &  -  \\\hline
$\Sigma^+\to \ell^+Z^0 $   &  $\ell^+ \ell^- Z^0 W^-  $   &
   $\ell^+ \ell^- Z^0 Z^0  $     &   $\ell^+ \ell^+ Z^0 W^-  $     &
  $\ell^+ \ell^- Z^0 W^+  $ \\
$ (0.06)$   &  $(0.04) $   &
   $ (0.004) $     &   $ (0.02) $     &
  $ (0.02) $ \\\hline
$\Sigma^0\to \ell^-W^+ $   &  -    &   $\ell^- \ell^-W^+ Z^0  $     &
-     &  - \\
$(0.30)$   &  -    &   $(0.02) $     &   -     &  - \\\hline
$\Sigma^0\to \ell^+W^- $   &  -    &   $\ell^+ \ell^-W^- Z^0  $     &
-     &  - \\
$(0.30)$   &  -    &   $(0.02) $     &   -     &  - \\\hline
\end{tabular}
\caption{\footnotesize Decays of exotic leptons to SM charged leptons,
including multi-lepton and same-sign dilepton events, together with
their branching ratios (restricted to $l=e, \mu$) and for $M_\Sigma =
400\,{\rm GeV}$.}
\label{events}
\end{table}

\newpage

\section{Possible role as dark matter}

The fermionic quintuplet of this paper was selected previously as a viable MDM candidate in \cite{Cirelli:2005uq}. To employ it as a seesaw mediator requires additional scalar multiplets, what makes the neutral component of the quintuplet unstable. If we employ a $Z_2$ symmetry under which $\Sigma \to - \Sigma$, $\Phi \to - \Phi$ and all SM fields are unchanged, the lightest component of the $\Sigma$ and $\Phi$ multiplets is again a DM candidate. The quartic $\lambda_5$ term in the scalar potential is still allowed so that neutrinos acquire radiatively generated masses given by Eq.~(\ref{loop}).

If $\Sigma^0$ is the DM particle, its mass is fixed by the relic abundance to the value $M_\Sigma \approx 10\ \rm{TeV}$ in \cite{Cirelli:2005uq}. The choice $\lambda_5=10^{-7}$ gives enough suppression to have small neutrino masses with large Yukawas, $Y \sim 0.1$. In this part of the parameter space the model could have interesting LFV effects like in \cite{Cai:2011qr}.

The neutral component of the scalar multiplet $\Phi$ could also be a DM particle, despite its tree level coupling to the $Z$ boson. Similar to the inert doublet model, the $\lambda_5$ term in the scalar potential in Eq.~(\ref{scalarpot}) splits the real and imaginary parts of the $\Phi^0$ field. If the mass splitting is large enough, the inelastic scattering through the $Z$ boson exchange is kinematically forbidden in direct detection experiments. In this case the new states could be within reach of the LHC. To ensure a large enough mass splitting, the coupling $\lambda_5$ cannot be too small. Accordingly, from Eq.~(\ref{loop}), it follows that the Yukawa couplings have to be smaller, suppressing the LFV effects.

\section{Conclusion}

In this account we propose a TeV-scale model for neutrino masses based only on the gauge symmetry and the renormalizability of the SM. The model employs fermion quintuplets with zero hypercharge, which in isolation correspond
to cosmological minimal dark matter candidate. Such TeV-scale fermions in conjunction with the scalar quadruplet
can generate neutrino masses both by the tree-level contributions in Fig.~\ref{dim9op} and the loop level contributions in Fig.~\ref{oneloop}. Also, such states are expected to be abundantly produced at the LHC, and the distinctive signatures could come from doubly-charged components
of the fermionic quintuplets. For $5\ \rm{fb^{-1}}$ of already accumulated integrated luminosity at the LHC ($\sqrt{s}=7$ TeV), there could be  $\sim 500$ produced doubly-charged $\Sigma^{++}$ or $\overline{\Sigma^{++}}$ fermions, with mass $M_\Sigma=\rm{400\ GeV}$. The decays of these states to SM charged leptons have an interesting multilepton signature. There are, in addition, same-sign dilepton events displayed in Table 2. These events have a negligible SM background, as demonstrated \cite{Mukhopadhyay:2011xs,delAguila_Aguilar-Saavedra} in generic new physics scenario with lepton number violation. 
In particular, the detailed studies of fermionic triplets from type III seesaw scenario \cite{Li:2009mw,delAguila:2008cj} apply to our case
of fermionic quintuplets. The multilepton events listed in our Table 2 are in direct correspondence with the items
from type III seesaw given in Table 13 of Ref. \cite{delAguila:2008cj}, to which  also the analysis in \cite{Li:2009mw} refers.
The signals which are good for the discovery correspond to a relatively high signal rate and small SM background,
which is calculated by MadGraph \cite{Alwall:2011uj}.
In order to compare the signal and SM background cross sections, we  assume a particular choice of parameters leading to the branching ratios given in Table 2, restricted to $l={e, \mu}$ leptons in the final states.
In this respect we distinguish four classes of events containing $\Sigma^+$ decaying to $e^+$ or $\mu^+$ lepton and $Z^0 \to (\ell^+ \ell^-, q \bar{q})$ resonance to help in $\Sigma^+$ identification:
\begin{displaymath}
(i) \qquad p \, p \to \Sigma^+  \, \overline{\Sigma^+} \to (\ell^+Z^0) \, (\ell^-Z^0) \;,
\end{displaymath}
which has too small of a cross section (0.03 fb with respect to the SM background of 0.6 fb) at 7 TeV LHC;
\begin{displaymath}
(ii) \qquad p \, p \to \Sigma^+  \, \overline{\Sigma^0} \to (\ell^+Z^0) \, (\ell^-W^+) \;,
\end{displaymath}
which has a cross section of 0.7 fb, comparable to the SM background of 0.8 fb;
\begin{displaymath}
(iii) \qquad p \, p \to \Sigma^+  \, \overline{\Sigma^0} \to (\ell^+Z^0) \, (\ell^+W^-) \;,
\end{displaymath}
the LNV event having 0.7 fb with the same-sign dilepton state, which is nonexistent in the SM and thus
devoid of the SM background;
\begin{displaymath}
(iv) \qquad p \, p \to \Sigma^{++}  \, \overline{\Sigma^+} \to (\ell^+W^+) \, (\ell^-Z^0) \;,
\end{displaymath}
having a relatively high signal rate (1.1 fb with respect to SM background of 0.8 fb).
The classes $(iii)$ and $(iv)$ lead to signals elaborately detailed in \cite{Li:2009mw,delAguila:2008cj}, which 
can be rescaled to infer on the LHC discovery reach.

The model of neutrino mass generation presented here represents a generalization of the fermion triplet seesaw mediator from the type III seesaw model to a hypercharge zero quintuplet, which we postponed at the time of elaborating first its nonzero-hypercharge variant in  \cite{Picek:2009is}.
The hypercharge zero quintuplet has been proposed subsequently in \cite{Liao:2010cc}. One can question the possibility to distinguish between the present hypercharge zero quintuplet with a doubly-charged component and previously considered nonzero hypercharge
states \cite{Picek:2009is,Kumericki:2011hf}, which provide spectacular falsifiable
triply-charged signatures. Consequently, a nonobservation of the latter  would put in a
forefront the hypercharge zero quintuplets, which provide their neutral components as potential MDM candidates.
However, a blow to the stability of such DM comes from the presence of additional fields
needed to generate the described small neutrino masses. As usual, the stability can be restored by introducing the discrete symmetry under which the new states are odd, in which case neutrinos get mass only from radiative contributions.

\subsubsection*{Acknowledgements}

This work is supported by the Croatian Ministry  of Science, Education, and Sports under Contract No. 119-0982930-1016.

\end{document}